\documentclass[english,prd,twocolumn,superscriptaddress,nofootinbib,preprintnumbers]{revtex4}
\usepackage[latin1]{inputenc}
\usepackage{graphicx}
\usepackage{amssymb}

\makeatletter

\providecommand{\tabularnewline}{\\}

\makeatletter

\makeatletter

\makeatletter

\usepackage{amsfonts}

\usepackage{dcolumn}

\usepackage{hyperref}

\def\be{\begin{equation}}
\def\ee{\end{equation}}
\def\ba{\begin{eqnarray}}
\def\ea{\end{eqnarray}}
\def\bs{\begin{subequations}}
\def\es{\end{subequations}}

\newcommand{\rd}{{\rm d}}
\newcommand{\rr}{{\rm rad}}

\usepackage{color}

\makeatother

\makeatother

\makeatother

\usepackage{babel}
\makeatother
\begin{document}

\title{Phantom crossing, equation-of-state singularities, \\
 and local gravity constraints in $f(R)$ models}

\author{Luca Amendola}

\affiliation{INAF/Osservatorio Astronomico di Roma, Via Frascati 33\\
 00040 Monte Porzio Catone (Roma), Italy}

\author{Shinji Tsujikawa}

\affiliation{Department of Physics, Gunma National College of Technology, Gunma
371-8530, Japan}

\begin{abstract}
We identify the class of $f(R)$ dark energy models which have a viable
cosmology, i.e. a matter dominated epoch followed by a late-time acceleration.
The deviation from a $\Lambda$CDM model ($f=R-\Lambda$) is quantified
by the function $m=Rf_{,RR}/f_{,R}$. The matter epoch corresponds
to $m(r=-1)\simeq+0$ (where $r=-Rf_{,R}/f$) while the accelerated
attractor exists in the region $0\le m<1$. We find that the equation
of state $w_{{\rm DE}}$ of all such {}``viable'' $f(R)$ models
exhibits two features: $w_{{\rm DE}}$ diverges at some redshift $z_{c}$
and crosses the cosmological constant boundary ({}``phantom crossing'')
at a redshift $z_{b}$ smaller than $z_{c}$. Using the observational
data of Supernova Ia and Cosmic Microwave Background, we obtain the
constraint $m<{\cal O}(0.1)$ and we find that the phantom crossing
could occur at $z_{b}\gtrsim1$, i.e. within reach of observations.
If we add local gravity constraints, the bound on $m$ becomes very
stringent, with $m$ several orders of magnitude smaller than unity
in the region whose density is much larger than the present cosmological
density. The representative models that satisfy both cosmological
and local gravity constraints take the asymptotic form $m(r)=C(-r-1)^{p}$
with $p>1$ as $r$ approaches $-1$. 
\end{abstract}

\date{\today}

\maketitle

\section{Introduction}

Recent observations have continuously confirmed that about 70\% of
the present energy density of the universe consists of dark energy
(DE) that leads to an accelerated expansion \cite{review}. The simplest
DE scenario consistent with observations is the $\Lambda$CDM model
in which DE is identified as a cosmological constant $\Lambda$. There
are many other attempts to explain the origin of DE, which can be
broadly classified into two classes. The first class consists of modified
gravity models in which gravity is modified from Einstein theory,
whereas the second class corresponds to introducing a more or less
exotic form of matter (such as a scalar field \cite{quin}) to explain
the late-time acceleration. One of the main focus of current research
is to find a departure from the $\Lambda$CDM model by confronting
dynamical DE models with observations.

In this paper we shall study cosmological and local gravity constraints
on a class of modified gravity DE models \cite{fR} whose action is
a general function $f(R)$ in terms of a Ricci scalar $R$, i.e.,
\begin{equation}
S=\int{\rm d}^{4}x\sqrt{-g}\left[\frac{1}{2\kappa^{2}}f(R)+{\mathcal{L}}_{{\rm m}}+{\mathcal{L}}_{{\rm rad}}\right]\,,\label{action}\end{equation}
 where $\kappa^{2}=8\pi G=1/M_{{\rm pl}}^{2}$ while $G$ is a bare
gravitational constant and $M_{{\rm pl}}$ is a reduced Planck mass
(see also Refs.~\cite{star,Capo,early}). Here ${\mathcal{L}}_{{\rm m}}$
and ${\mathcal{L}}_{{\rm rad}}$ are the Lagrangian densities of dust-like
matter and radiation, respectively. Throughout this paper we shall
focus on the metric-variational approach. See Refs.~\cite{Pala}
for the Palatini formalism of $f(R)$ DE models.

In Ref.~\cite{APT} it was shown that the models of the types $f(R)=R-\alpha/R^{n}$
($n>0$) and $f(R)=\alpha R^{n}$ ($n\neq1$) do not possess a standard
matter epoch because of a large coupling between gravity and dark
matter, even though a late-time acceleration can be realized (see
also Refs.~\cite{others,Song,Li}). Extending the analysis to general
$f(R)$ cases, the paper \cite{AGPT} has recently clarified the conditions
under which $f(R)$ DE models have a matter era followed by an accelerated
expansion. However this does not necessarily mean that the models
satisfying the conditions derived in \cite{AGPT} can be consistent
with observations. In this paper we constrain $f(R)$ models that
have a matter epoch prior to the acceleration from the observational
data such as Supernova Ia (SNIa) and the sound horizon of Cosmic Microwave
Background (CMB).

The deviation from the $\Lambda$CDM model is quantified by a variable
\begin{eqnarray}
m=\frac{Rf_{,RR}}{f_{,R}}\,,\label{mdef}\end{eqnarray}
 where $f_{,R}\equiv{\rm d}f/{\rm d}R$ and $f_{,RR}\equiv{\rm d}^{2}f/{\rm d}R^{2}$.
Note that the $\Lambda$CDM model corresponds to $m=0$. As we will
show in this paper, the quantity $m$ can be constrained as $m<{\cal O}(0.1)$
throughout the matter and accelerated epochs from CMB and SNIa data.
This limit still allows for interesting deviations from $\Lambda$CDM,
in particular for the possibility to observe a phantom crossing and
a singularity at low redshifts in the equation of state of DE. However,
as we will show, local gravity constraints (LGC) give a tight bound
for $m$ very much smaller than unity in a region whose density is
much larger than the present cosmological density. Although this tight
constraint excludes most $f(R)$ models (or renders them indistinguishable
from the $\Lambda$CDM model), those proposed recently
by Hu \& Sawicki \cite{Hu07} and Starobinsky \cite{Starobinsky07}, 
which appeared after the initial submission of this article, 
are still viable and can be distinguished from $\Lambda$CDM.

\section{Viable cosmological trajectories}

In the flat Friedmann-Robertson-Walker background with a scale factor
$a$, the evolution equations in the metric-variational approach are
given by \begin{eqnarray}
3FH^{2} & = & \kappa^{2}\,(\rho_{{\rm m}}+\rho_{{\rm rad}})+(FR-f)/2-3H\dot{F},\label{E1}\\
-2F\dot{H} & = & \kappa^{2}\left[\rho_{{\rm m}}+(4/3)\rho_{{\rm rad}}\right]+\ddot{F}-H\dot{F},\label{E2}\end{eqnarray}
 where $F=\partial f/\partial R$, $H=\dot{a}/a$, $R=6(2H^{2}+\dot{H})$,
and a dot denotes a derivative in terms of cosmic time $t$. Note
that we study the dynamics in the positive $F$ branch. The energy
densities of a non-relativistic matter and radiation satisfy the equations
\begin{eqnarray}
 &  & \dot{\rho}_{{\rm m}}+3H\rho_{{\rm m}}=0\,,\\
 &  & \dot{\rho}_{\rr}+4H\rho_{\rr}=0\,,\end{eqnarray}
 respectively.

In order to confront the models with SNIa observations, it is convenient
to write the equations as follows \begin{eqnarray}
3F_{0}H^{2} & = & \kappa^{2}(\rho_{{\rm DE}}+\rho_{{\rm m}}+\rho_{{\rm rad}})\,,\label{E10}\\
2F_{0}\dot{H} & = & -\kappa^{2}\left[\rho_{{\rm m}}+(4/3)\rho_{{\rm rad}}+\rho_{{\rm DE}}+p_{{\rm DE}}\right]\,,\label{E2a}\end{eqnarray}
 where \begin{eqnarray}
\kappa^{2}\rho_{{\rm DE}} & = & (1/2)(FR-f)-3H\dot{F}+3H^{2}(F_{0}-F)\,,\label{rhoDE}\\
\kappa^{2}p_{{\rm DE}} & = & \ddot{F}+2H\dot{F}-(1/2)(FR-f)\nonumber \\
 &  & -(2\dot{H}+3H^{2})(F_{0}-F)\,.\end{eqnarray}
 Here the subscript {}``0'' represents present values at the redshift
$z=0$. By defining $\rho_{{\rm DE}}$ and $p_{{\rm DE}}$ in the
above way, these satisfy the usual conservation equation \begin{eqnarray}
{\dot{\rho}_{{\rm DE}}}+3H(\rho_{{\rm DE}}+p_{{\rm DE}})=0\,.\end{eqnarray}
 Then the DE equation of state (EOS) parameter, $w_{{\rm DE}}\equiv p_{{\rm DE}}/\rho_{{\rm DE}}$,
is directly related to the one obtained from observations \cite{efp}.

Introducing the following variables \begin{eqnarray*}
x_{1}=-\frac{\dot{F}}{HF},~x_{2}=-\frac{f}{6FH^{2}},~x_{3}=\frac{R}{6H^{2}},~x_{4}=\frac{\kappa^{2}\rho_{{\rm rad}}}{3FH^{2}},\end{eqnarray*}
 we obtain \cite{AGPT} \begin{eqnarray}
x_{1,N} & = & -1-x_{3}-3x_{2}+x_{1}^{2}-x_{1}x_{3}+x_{4}~,\label{N1}\\
x_{2,N} & = & \frac{x_{1}x_{3}}{m}-x_{2}(2x_{3}-4-x_{1})~,\label{N2}\\
x_{3,N} & = & -\frac{x_{1}x_{3}}{m}-2x_{3}(x_{3}-2)~,\label{N3}\\
x_{4,N} & = & -2x_{3}x_{4}+x_{1}\, x_{4}\,,\label{N4}\end{eqnarray}
 where $N=\ln a$, $x_{i,N}={\rm d}x_{i}/{\rm d}N$ and \begin{eqnarray}
m & = & \frac{\rd\log F}{\rd\log R}=\frac{Rf_{,RR}}{f_{,R}}\,,\label{mdef2}\\
r & = & -\frac{\rd\log f}{\rd\log R}=-\frac{Rf_{,R}}{f}=\frac{x_{3}}{x_{2}}\,.\label{ldef}\end{eqnarray}
 Deriving $R$ as a function of $r$ from Eq.~(\ref{ldef}), one
can express $m$ as a function of $r=x_{3}/x_{2}$ and close the above
system. Notice that defining $\Omega_{{\rm m}}\equiv\kappa^{2}\rho_{{\rm m}}/3FH^{2}$
one has \begin{eqnarray}
\Omega_{{\rm m}}=1-x_{1}-x_{2}-x_{3}-x_{4}\,.\end{eqnarray}

The DE equation of state is given by \begin{eqnarray}
w_{{\rm DE}}=\frac{p_{{\rm DE}}}{\rho_{{\rm DE}}}=-\frac{1}{3}\frac{2x_{3}-1+x_{4}y}{1-y(1-x_{1}-x_{2}-x_{3})}\,,\label{wDEx}\end{eqnarray}
 where $y\equiv F/F_{0}$. We note that the effective equation of
state of the system is given by $w_{{\rm eff}}=-(2x_{3}-1)/3$.

The analysis of the phase space has been performed in great detail
in \cite{AGPT}. Here we summarize the main results. In the absence
of radiation ($x_{4}=0$) we have six fixed points for the above system.
A matter epoch can be realized for $m\approx0$ and $r\approx-1$
on the critical point \begin{eqnarray*}
P_{M}:~(x_{1},x_{2},x_{3})=\left(\frac{3m}{1+m},-\frac{1+4m}{2(1+m)^{2}},\frac{1+4m}{2(1+m)}\right),\end{eqnarray*}
 which satisfies $w_{{\rm eff}}=-\frac{m}{1+m}$. If $m\approx+0$
and $m'(r)\equiv{\rm d}m/{\rm d}r>-1$ at $r\approx-1$ the matter
era corresponds to a saddle with a damped oscillation, whereas if
$m<0$ a prolonged matter period is not realized because the real
part of the eigenvalues of the critical point diverges (see also \cite{Song}).
Note that the radiation point also exists in the region with $m\approx0$,
which corresponds to a saddle \cite{AGPT}. Hence a viable cosmological
trajectory starts around the radiation point with $m\approx0$, which
is followed by the matter point with $m\approx0$.

When the trajectory passes through a standard matter era $P_{M}$
with $m\approx+0$, then one sees that $\Omega_{{\rm m}}\approx1$.
On the other hand, the future asymptotic value of $\Omega_{{\rm m}}$
is always zero if the acceleration occurs (see below). Therefore the
denominator of $w_{{\rm DE}}$, which in the absence of radiation
can be written as $1-F\Omega_{{\rm m}}/F_{0}$, goes from $\approx1$
in the future to $1-F/F_{0}$ in the deep matter era; this shows that
if $F$ increases toward the past $1-F/F_{0}$ crosses zero and becomes
negative and consequently $w_{{\rm DE}}$ passes necessarily through
a singularity. We will show in the next section that this is indeed
what happens.

\begin{figure}
\begin{centering}\includegraphics[width=3.4in,height=3.1in]{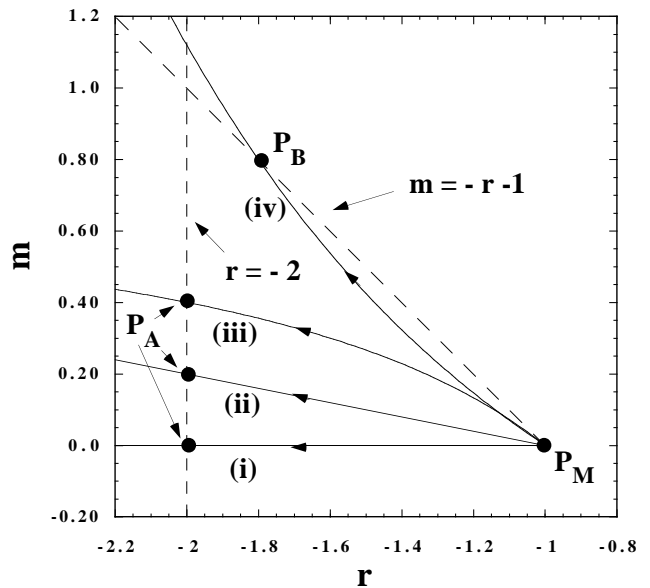} \par\end{centering}

\caption{\label{tra} 4 trajectories in the $(r,m)$ plane. Each trajectory
corresponds to (i) $\Lambda$CDM, (ii) $f(R)=(R^{b}-\Lambda)^{c}$,
(iii) $f(R)=R-\alpha R^{n}$ with $\alpha>0,0<n<1$, and (iv) $m(r)=-C(r+1)(r^{2}+ar+b)$.
Here $P_{M}$, $P_{A}$ and $P_{B}$ are matter, de-Sitter and non-phantom
accelerated points, respectively.}
\end{figure}

There are two stable fixed points leading to a late-time acceleration:
\begin{eqnarray}
{\rm (i)}~ & P_{A} & :~(x_{1},x_{2},x_{3})=(0,-1,2),\\
{\rm (ii)}~ & P_{B} & :~(x_{1},x_{2},x_{3})\nonumber \\
 &  & =\left(\frac{2(1-m)}{1+2m},\frac{1-4m}{m(1+2m)},-\frac{(1-4m)(1+m)}{m(1+2m)}\right).\nonumber \end{eqnarray}
 The effective EOS is given by $w_{{\rm eff}}=-1$ for $P_{A}$ and
$w_{{\rm eff}}=\frac{2-5m-6m^{2}}{3m(1+2m)}$ for $P_{B}$. The former
exists on the line $r=-2$ and is stable for $0<m\le1$. The latter
exists on the line $m(r)=-r-1$ as is the case for the point $P_{M}$.
There are several ranges of $m$ that lead to a stable acceleration,
but only for $(\sqrt{3}-1)/2<m<1$ and $m'(r)<-1$ one can have a
transition from the saddle matter era to the accelerated epoch (in
this case $w_{{\rm eff}}>-1$) (see discussions in \cite{AGPT}).
Therefore, we have only two qualitatively different viable cases:

\begin{itemize}
\item Models that link $P_{M}$ with $P_{A}$ (Class A), 
\item Models that link $P_{M}$ with $P_{B}$ (Class B). 
\end{itemize}
See Fig.~\ref{tra} for an illustration.

The cosmological dynamics of $f(R)$ models can be well understood
by considering $m(r)$ curves in the $(r,m)$ plane. The $\Lambda$CDM
model, $f(R)=R-\Lambda$, corresponds to $m=0$, in which case the
trajectory is a straight line from $P_{M}$: $(r,m)=(-1,0)$ to $P_{A}$:
$(r,m)=(-2,0)$. We now introduce three $f(R)$ toy models that represent
the two qualitatively different classes of cosmologies A and B. As
Class A we define a minimal generalization of the $\Lambda$CDM model
given by (model A1) \begin{eqnarray}
f(R)=(R^{b}-\Lambda)^{c}\,.\label{stra}\end{eqnarray}
 This is characterized by the straight line \begin{eqnarray}
m(r)=[(1-c)/c]r+b-1\,.\end{eqnarray}
 The existence of a saddle matter epoch requires the condition $c\ge1$
and $bc\approx1$. When $c=1$ the trajectory is given by $m=b-1$
and is parallel to the $\Lambda$CDM line ($m=0$). We also introduce
models of the type (model A2) \begin{eqnarray}
f(R)=R-\alpha R^{n}\,,\quad\alpha>0,~~0<n<1\,,\label{A2}\end{eqnarray}
which satisfy $m=n(1+r)/r$ and also fall into the Class A \cite{AGPT,Li}.
As representative of Class B, i.e., a cosmological evolution from
$P_{M}$ to $P_{B}$, we select the models (model B1) \begin{eqnarray}
m(r)=-C(r+1)(r^{2}+ar+b)\,,\end{eqnarray}
 where $C>0$. We require the conditions $m'(-1)=-C(1-a+b)>-1$ and
$m'(-2)=C(3a-b-8)<-1$ for the transition from the matter era to the
stable acceleration. In Fig.~\ref{tra} we plot four trajectories
corresponding to the various cases we presented above. These toy models
are meant only to exemplify the classes of viable cosmologies; it
is clear that there are an infinite number of $m(r)$ curves connecting
$P_{M}$ to $P_{A,B}$. However, most of our arguments below will
apply to any of these curves.

\section{Observational constraints}

For the viability of $f(R)$ models we require that they satisfy three
observational constraints: (i) CMB, (ii) SNIa and (iii) LGC. The first
one comes from the angular size of the sound horizon defined by \begin{eqnarray}
\Theta_{s}=\int_{z_{{\rm {dec}}}}^{\infty}\frac{c_{s}(z){\rm {d}}z}{H(z)}\,\Biggl/\int_{0}^{z_{{\rm {dec}}}}\frac{{\rm {d}}z}{H(z)}\,,\end{eqnarray}
 where $c_{s}^{2}(z)=1/[3(1+3\rho_{b}/4\rho_{\gamma})]$ is the adiabatic
baryon-photon sound speed and $z_{{\rm {dec}}}\simeq1089$ is the
redshift at the decoupling time. This quantity $\Theta_{s}$ is related
to the position of CMB acoustic peaks and has been constrained as
$\Theta_{s}=0.5946\pm0.0021$ deg from the WMAP 3year data \cite{Spergel}.
Since the eigenvalues of the Jacobian matrix for perturbations about
the matter point $P_{M}$ are given by $3(1+m'(r))$ and $-3/4\pm\sqrt{-1/m}$,
the matter epoch ($m\approx+0$) lasts for a long time as the tangent
$m'(-1)$ approaches $-1$. If $m'(-1)$ is close to $-1$ it is generally
difficult to satisfy the CMB constraint.

We have integrated the autonomous equations to obtain the present
matter density $\Omega_{{\rm m}}^{(0)}=0.28$. Then we solved back
the equations toward the past to get $\Theta_{s}=0.5946\pm0.0021$.
This is a trial and error procedure that is done by changing initial
conditions in the radiation era. If the present radiation density
satisfies the condition $5.0\times10^{-5}<\Omega_{{\rm rad}}^{(0)}<2.5\times10^{-4}$
together with the sound horizon constraint, we conclude that the models
pass the CMB test. Then the straight line model A1 given in (\ref{stra}),
for example, is constrained to $c<3$ and $m<0.282$. For larger $c$
the contribution of dark energy is significant even in the matter-dominated
epoch, thus incompatible with the CMB constraint. This comes from
the fact that as $m$ deviates from 0 the coupling between dark energy
and dark matter becomes significant.

In what follows we shall consider the SNIa constraint under the situation
where the CMB constraint is satisfied. The viable cosmological trajectories
are restricted to be in the range $m>0$ and $r<0$. Since we are
considering the case of positive $F$, this translates into the conditions
$RF_{,R}>0$ and $R/f>0$. If $R$ changes sign, both $m$ and $r$
change signs simultaneously. Then it is sufficient to consider positive
$R$, which gives $F_{,R}>0$. This translates into the condition
$\dot{F}<0$ when $R$ decreases in time, which means that the variable
$F$ increases toward the past. Hence the DE equation of state exhibits
a divergence at some redshift $z_{c}>0$ because $\rho_{{\rm DE}}$
changes sign. Since $p_{{\rm DE}}$ is negative provided $x_{3}\ge1/2$
(in which case the radiation, matter and accelerated epochs are included),
we have $w_{{\rm DE}}<0$ for $z<z_{c}$ and $w_{{\rm DE}}>0$ for
$z>z_{c}$ together with the singularity $w_{{\rm DE}}\to-\infty$
as $z\to-z_{c}$ and $w_{{\rm DE}}\to+\infty$ as $z\to+z_{c}$. %
\footnote{We note that a divergence of $w_{{\rm DE}}$ can occur for a Dvali-Gabadadze-Porrati
braneworld model \cite{Alam}.%
} This shows an interesting feature of $f(R)$ models: whenever they
are cosmologically acceptable they exhibit a singularity in $w_{{\rm DE}}$
at some epoch in the past and, as a consequence, a crossing of the
{}``phantom boundary'' $w_{{\rm DE}}=-1$. Note that the models
with $F_{,R}<0$ (such as $f(R)=R-\alpha/R^{n}$, $n>0$) do not exhibit
such peculiar behavior, but they are not cosmologically viable \cite{APT}.

\begin{figure}
\begin{centering}\includegraphics[width=3.3in,height=3in]{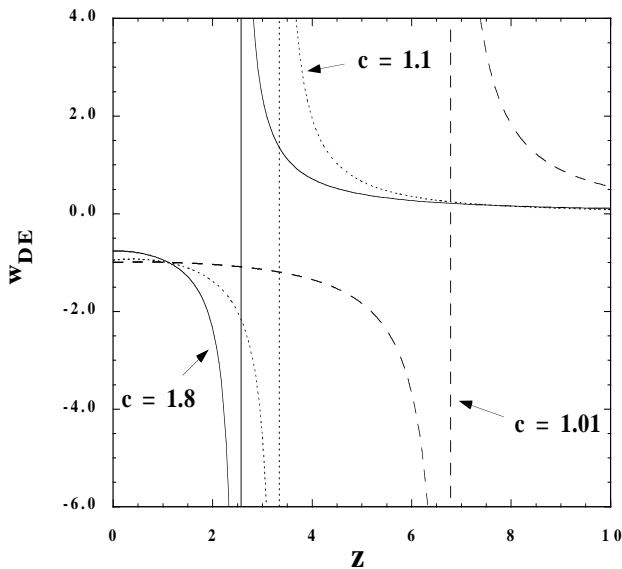} \par\end{centering}

\caption{\label{w} Evolution of the DE equation of state $w_{{\rm DE}}$
for the model $f(R)=(R^{1/c}-\Lambda)^{c}$ with parameters $c=1.01,1.1,1.8$.
As $c$ approaches 1, the critical value $z_{c}$ gets larger. In
the limit $c\to1$ ($\Lambda$CDM model) we have $z_{c}\to\infty$. }
\end{figure}

In Fig.~\ref{w} we plot the evolution of $w_{{\rm DE}}$ for the
model (\ref{stra}) with $bc=1$ for three different values of $c$.
The $\Lambda$CDM model ($c=1$) corresponds to $z_{c}\to\infty$.
As $c$ deviates from 1, the critical redshift $z_{c}$ gets smaller
together with the increase of the present value of $w_{{\rm DE}}$
departing from $-1$. The phantom crossing ($w_{{\rm DE}}=-1$) is
realized at the redshift $z_{b}$ smaller than $z_{c}$. It is worth
pointing out that the phantom crossing occurs from the region $w_{{\rm DE}}<-1$
to the region $w_{{\rm DE}}>-1$, which is different from quintom
models of DE \cite{quintom}. In Table I we show the values $z_{b}$,
$z_{c}$, $w_{{\rm DE}}(z=0)$ and $m(z=0)$ with several different
choices of $c$. We find that $z_{b}$ is generally close to unity
(unless $w_{{\rm DE}}$ today is extremely close to $-1$), which
is within the observational range of SNIa. This interesting feature
could be employed to discriminate $f(R)$ modified gravity from other
dark energy models. The divergence of $w_{{\rm DE}}$ occurs at a
redshift larger than $z=2$ (if one also imposes the CMB constraints),
so this is outside of the current observational range of SNIa.

%
\begin{table}[t]
\begin{tabular}{|c|c|c|c|c|}
\hline 
$c$ &
$z_{b}$ &
$z_{c}$ &
$w_{{\rm DE}}(z=0)$ &
$m(z=0)$\tabularnewline
\hline 
$1.01$ &
$1.20$ &
$6.77$ &
$-0.996$ &
$0.008$ \tabularnewline
\hline 
$1.1$ &
$1.09$ &
$3.33$ &
$-0.952$ &
$0.076$\tabularnewline
\hline 
$1.5$ &
$1.05$ &
$2.55$ &
$-0.818$ &
$0.222$\tabularnewline
\hline 
$1.8$ &
$1.12$ &
$2.52$ &
$-0.766$ &
$0.256$ \tabularnewline
\hline 
$2.3$ &
$1.24$ &
$2.61$ &
$-0.705$ &
$0.276$ \tabularnewline
\hline
\end{tabular}

\caption[table1]{\label{table1} The values of $z_{b}$, $z_{c}$, $w_{{\rm DE}}(z=0)$
and $m(z=0)$ for the model $f(R)=(R^{1/c}-\Lambda)^{c}$. The present
epoch corresponds to $\Omega_{{\rm m}}^{(0)}=0.28$.}
\end{table}

\begin{figure}
\begin{centering}\includegraphics[width=3.3in,height=3in]{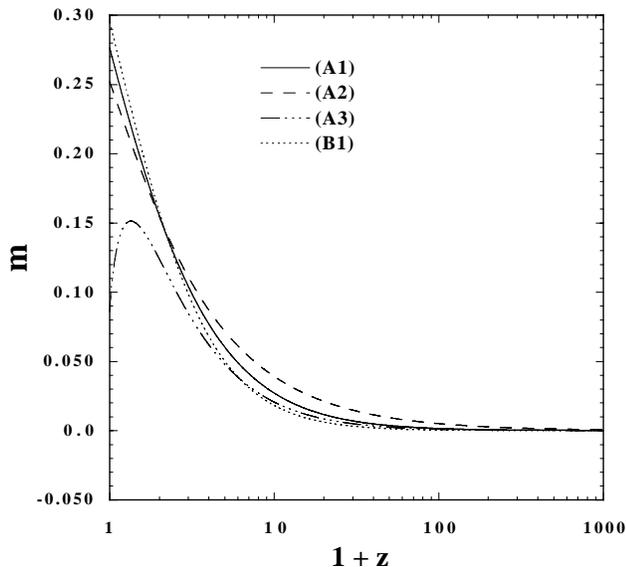} \par\end{centering}

\caption{\label{mz} Evolution of the variable $m$ in terms of $z$ for the
models: (A1) $f=(R^{1/c}-\Lambda)^{c}$, (A2) $f=R-\alpha R^{n}~(0<n<1)$,
(A3) $m=-C(r+1)(r+2.1)$ and (B1) $m=-C(r+1)(r^{2}+r+1)$. Each curve
shows the maximal $m$ that still satisfies SNIa and CMB constraints. }
\end{figure}

We can also use the criterion $w_{{\rm DE}}(z=0)<-0.7$ for the compatibility
with the SNIa data \cite{Astier}. Then in the model A1 we obtain
the constraint $c<2.3$ and $m<0.276$, which is slightly stronger
than the one obtained by the CMB. We have also carried out a numerical
analysis for the other $f(R)$ models A2 and B1 described in the previous
section. In Fig.~\ref{mz} the evolution of $m$ is plotted as a
function of $z$ in the marginal cases satisfying both the CMB and
SNIa requirements. The constraints on $m$ for the models (A1) $f=(R^{1/c}-\Lambda)^{c}$
; (A2) $f=R-\alpha R^{n}~(0<n<1)$, and (B1) $m=-C(r+1)(r^{2}+r+1)$
are similar at all redshifts. Adopting instead a further model (A3)
$m=-C(r+1)(r+2.1)$ (which also belongs to the class A) gives a tighter
constraint near the present epoch. This is simply due to the fact
that $m$ decreases as $r$ approaches $-2$ after having a maximum
value of $m$ at $r=-1.55$ in the $(r,m)$ plane. The difference
between class A and B models is not significant provided the point
$P_{B}$ exists in the large $m$ region close to 1 to ensure sufficient
acceleration. In Table \ref{table2} we summarize the maximum values
of $m$ and the allowed model parameters. As one can see, the parameter
$m$ is constrained to be $m<0.1$-$0.3$ in all models we have considered.
We have also checked that the slopes of the EOS, $|{\rm d}w_{{\rm DE}}/{\rm d}z|$,
are smaller than $\approx\,$0.1 at present epoch and do not provide
better constraints than the ones obtained from the criterion $w_{{\rm DE}}(z=0)<-0.7$.

\begin{table}[t]
\begin{tabular}{|c|c|c|}
\hline 
${\rm Model}$ &
${\rm Constraints}$ &
$z_{c}$ \tabularnewline
\hline 
$f=(R^{1/c}-\Lambda)^{c}$ &
$m<0.276,~c<2.3$ &
$2.61$ \tabularnewline
\hline 
$f=R-\alpha R^{n}~(0<n<1)$ &
$m<0.252,~n<0.7$ &
$2.87$ \tabularnewline
\hline 
$m=-C(r+1)(r+2.1)$ &
$m<0.151,~C<0.5$ &
$2.95$ \tabularnewline
\hline 
$m=-C(r+1)(r^{2}+r+1)$ &
$m<0.295,~1/3<C<0.45$ &
$2.40$ \tabularnewline
\hline
\end{tabular}

\caption[table2]{\label{table2} The constraint on the parameter $m$ for several
$f(R)$ models coming from the SNIa constraint $w_{{\rm DE}}(z=0)<-0.7$
and the CMB. We also show the value $z_{c}$ corresponding to the
maximum allowed $m$. }
\end{table}

As we have seen, the cosmological observations require $m$ to lie
below 0.1-0.3 at all redshifts (up to the radiation epoch). For these
values, the models display a phantom crossing and an equation of state
singularity at a redshift of a few, making such $f(R)$ theories quite
intriguing from the observational point of view, especially since
some SN analysis finds some evidence for a similar crossing (see e.g.
\cite{nesper}).

Let us now include the constraints from local gravity experiments.
The Newtonian effective gravitational constant can be obtained under
a weak-field approximation by considering a spherically symmetric
body with a mass $M_{\odot}$, constant density $\rho$ and a radius
$r_{\odot}$ and a vanishing density ($\rho=0$) outside the body.
Using a linear perturbation theory in the Minkowski background with
a perturbation $h_{\mu\nu}$ and decomposing the function $F$ into
background and perturbation parts ($F=F_{0}+\delta F$), the effective
gravitational potential at a distance $\ell$ from the center of the
body is \cite{amelnl,Olmo,CSE,Navarro} 
\begin{eqnarray}
G_{{\rm eff}}=\frac{G}{F_{0}}
\left(1+\frac{1}{3}e^{-M \ell}\right)\,,
\label{effectiveGM}
\end{eqnarray}
where the mass $M$ is given by 
\begin{eqnarray}
M^{2}=\frac{R}{3}\left(\frac{f_{,R}}{Rf_{,RR}}-1\right)
=\frac{R}{3m}(1-m)\,,\label{Massdef}
\end{eqnarray}
If $M^{2}<0$ the Yukawa correction $e^{-M\ell}$ in Eq.~(\ref{effectiveGM})
is replaced by an oscillating function $\cos(|M|\ell)$, but this
case is excluded experimentally. Hence the mass squared is required
to be positive.

As emphasized in Ref.~\cite{CSE}, the expression (\ref{effectiveGM})
is valid in the regime in which the linear approximation $|\delta R|\ll R_{0}$
holds (here $\delta R$ is a perturbation in $R$ with $R_{0}$ being
a background value). The condition for the validity of the linear
approximation is given in Eq.~(33) of Ref.~\cite{CSE}, which is
equivalent to 
\begin{eqnarray}
m(R_{0})\gg\Phi_{c}\,,\label{mcon}
\end{eqnarray}
where $m$ is defined in Eq.~(\ref{mdef}) and $\Phi_{c}=GM_{\odot}/r_{\odot}$
is the gravitational potential at the surface of the body. Since the
mass squared $M^{2}$ is estimated as $M^{2}\simeq R/3m$ for $m\ll1$,
the condition (\ref{mcon}) tends to be violated for large $M$. 
If we require $M\ell \gg1$ with $\ell$ being the scale of the experiment 
(e.g. a scale of 1\,mm for laboratory constraints or a scale of 1 AU for
solar system constraints), we obtain 
\begin{equation}
m\ll 8\pi G_N \ell^{2}\rho\,, \label{eq:mmeas}
\end{equation}
where $G_N \equiv G/F_{0}$ is the gravitational constant measured
at scales much larger than $\ell$.
Note that we used the relation 
$R \approx\kappa^{2}\rho/F_{0}$.
This constraint is extremely stringent:
assuming e.g. $\ell=1$\,AU and $\rho=10^{-23}$\,g/cm$^3$ 
(corresponding to the average density in the solar system) one gets $m\ll10^{-23}$.
Similar (or stronger) values are obtained for other experimental settings
on the Earth or near other solar-system bodies. 
Combining Eqs.~(\ref{mcon}) and (\ref{eq:mmeas})
we see that the condition for the applicability
at a distance $\ell$ from the center of a spherical structure 
of the linear perturbation approach is that 
\begin{equation}
\ell^{3}\rho(\ell)\gg\int_{0}^{\ell}
\rho(\ell')\ell'^{2}{\rm d}\ell'\,,
\end{equation}
which, for most astrophysical bodies, is actually violated. 
In particular, since $\Phi_{c}\approx10^{-9} \sim 10^{-6}$ 
for the Earth or the Sun or other planetary bodies, 
we find that Eq.~(\ref{mcon})
is actually grossly violated for local gravity experiments and we
need to consider the non-linear regime.

When the mass $M$ is heavy, the system enters a non-linear stage
in which a thin-shell develops inside the body through a chameleon
mechanism \cite{kw}. In order to consider the chameleon effect in
$f(R)$ gravity, it is convenient to transform to the Einstein frame
by a conformal transformation. Introducing a scalar field $\phi$
as $F=\exp(\sqrt{2/3}\kappa\phi)$, the potential in the Einstein
frame is given by $V=(RF-f)/2\kappa^{2}F^{2}$ with a constant coupling
$\beta=-1/\sqrt{6}$ between the matter and the field $\phi$ \cite{Maeda,APT}.

In a spherically symmetric setting with an energy density $\rho$,
we obtain the following equation for the field $\phi$ \cite{kw,Faul}
\begin{eqnarray}
\frac{{\rm d}^{2}\phi}{{\rm d}\tilde{\ell}^{2}}
+\frac{2}{\tilde{\ell}}\frac{{\rm d}\phi}{{\rm d}\tilde{\ell}}
=\frac{{\rm d}V_{{\rm eff}}}{{\rm d}\phi}\,,\label{dreq}\end{eqnarray}
where $\tilde{\ell}$ is the distance from the center of symmetry in
the Einstein frame and \begin{eqnarray}
V_{{\rm eff}}(\phi)=V(\phi)+e^{\beta\kappa\phi}\rho^{*}\,.\label{Veff}\end{eqnarray}
The energy density $\rho^{*}$ is defined by $\rho^{*}\equiv e^{3\beta\kappa\phi}\rho$,
which is conserved in the Einstein frame \cite{kw}.

Let us consider a spherically symmetric body with an energy density
$\rho^{*}=\rho_{A}^{*}$ inside the body ($\tilde{\ell}<\tilde{r}_{\odot}$)
and an energy density $\rho^{*}=\rho_{B}^{*}\ll\rho_{A}^{*}$ outside
the body ($\tilde{\ell}>\tilde{r}_{\odot}$). Then the effective potential
(\ref{Veff}) has two minima at $\phi=\phi_{A}$ and $\phi=\phi_{B}$
satisfying the relations $V_{,\phi}(\phi_{A})+\beta\kappa e^{\beta\kappa\phi_{A}}\rho_{A}^{*}=0$
and $V_{,\phi}(\phi_{B})+\beta\kappa e^{\beta\kappa\phi_{B}}\rho_{B}^{*}=0$,
respectively. The effective masses at the potential minima are defined
by $m_{A}^{2}\equiv V_{{\rm eff}}''(\phi_{A})$ and 
$m_{B}^{2}\equiv V_{{\rm eff}}''(\phi_{B})$,
where the mass $m_{A}$ is much heavier than the mass $m_{B}$.

When a body has a thin shell, the solution to Eq.~(\ref{dreq}) in
the region $\tilde{\ell}>\tilde{r}_{\odot}$ is approximately 
given by \cite{kw,Navarro,Faul,TUT} 
\begin{eqnarray}
\phi(\tilde{\ell})\simeq-
\frac{\beta_{{\rm eff}}}{4\pi M_{{\rm pl}}}
\frac{M_{\odot}e^{-m_{B}(\tilde{\ell}-\tilde{r}_{\odot})}}
{\tilde{\ell}}+\phi_{B}\,,\label{phir}\end{eqnarray}
where $M_{\odot}=4\pi r_{\odot}^{3}\rho_{A}/3
=4\pi\tilde{r}_{\odot}^{3}\rho_{A}^{*}/3$,
\begin{eqnarray}
\beta_{{\rm eff}}=3\beta\frac{\Delta\tilde{r}_{\odot}}
{\tilde{r}_{\odot}}\,,\quad
\frac{\Delta\tilde{r}_{\odot}}{\tilde{r}_{\odot}}
=-\frac{\phi_{B}-\phi_{A}}{\sqrt{6}M_{{\rm pl}}
\Phi_{\odot}}\,,\label{phiBA}
\end{eqnarray}
and $\Phi_{\odot}=GM_{\odot}/\tilde{r}_{\odot}$. As long as the thin-shell
condition $\Delta\tilde{r}_{\odot}/\tilde{r}_{\odot}\ll1$ is satisfied,
the effective coupling $|\beta_{{\rm eff}}|$ becomes much smaller
than unity. In this case the models can be consistent with the results
of solar system experiments as well as equivalence principle experiments.
For example, in the case of two identical bodies with mass $M_c$, 
the potential energy associated with the fifth force between the 
bodies is given by $U(\ell)=2\beta_{\rm eff}^2 
(GM_c^2/\ell)\,e^{-m_B\ell}$ \cite{kw}.
The laboratory experiment constraints, $2\beta_{\rm eff}^2<10^{-3}$, 
are satisfied for $\Delta\tilde{r}_{\odot}/\tilde{r}_{\odot}\ll1$.

In order to understand the condition under which the body has a thin
shell, let us consider the model (\ref{A2}) with $\alpha=\lambda R_{c}^{1-n}$.
Note that $R_{c}$ is not much different from the order of the present
cosmological constant. In this case the Ricci scalar $R_{1}$ at the
de-Sitter point $P_{A}$ satisfies the relation $\lambda=(R_{1}/R_{c})^{1-n}/(2-n)$.
In the region with a high density satisfying the relation $R\gg R_{c}$,
the field $\phi_{B}$ and the parameter $m(R)$ are approximately
given by \begin{eqnarray}
\phi_{B} & \simeq & -\frac{\sqrt{6}}{2}\lambda n\left(\frac{R_{c}}{\kappa^{2}\rho_{B}}\right)^{1-n}M_{{\rm pl}}\,,\label{phiB}\\
m(R) & \simeq & \lambda n(1-n)\left(\frac{R_{c}}{R}\right)^{1-n}\,.\end{eqnarray}
Using the fact that the Ricci scalar $R_{B}$ in the region B is approximated
by $R_{B}\simeq\kappa^{2}\rho_{B}$, we find that the thin-shell parameter
is \begin{eqnarray}
\frac{\Delta\tilde{r}_{\odot}}
{\tilde{r}_{\odot}}\simeq\frac{1}{2(1-n)}
\frac{m(R_{B})}{\Phi_{\odot}}\,.\label{delr}\end{eqnarray}
This shows that, unless $n$ is very close to 1, the thin-shell condition
$\Delta\tilde{r}_{\odot}/\tilde{r}_{\odot}\ll1$ holds for 
\begin{eqnarray}
m(R_{B})\ll\Phi_{\odot}\,,\label{mcond}\end{eqnarray}
which is opposite to the condition (\ref{mcon}) for the validity
of the linear perturbation theory. When the non-linearity becomes
important the body has a thin-shell. Since $\Phi_{\odot}\sim10^{-6}$
and $10^{-9}$ for the Sun and the Earth respectively,
the condition (\ref{mcond}) shows that the parameter $m$ is very
much smaller than unity in a high-density region where local gravity
experiments are carried out ($R_{B}\gg R_{c}$).

The current tightest constraint on the post-Newtonian parameter $\gamma$
in solar-system tests comes from Cassini tracking, which gives $|\gamma-1|<2.3\times10^{-5}$
\cite{Will}, This translates into the bound \cite{Faul} \begin{eqnarray}
\frac{\Delta\tilde{r}_{\odot}}{\tilde{r}_{\odot}}<1.15\times10^{-5}\,.\end{eqnarray}
Using this bound for Eq.~(\ref{delr}) with the value $\Phi_{\odot}\simeq2.12\times10^{-6}$
of the Sun, we obtain \begin{eqnarray}
\frac{n}{2-n}\left(\frac{\rho_{1}}{\rho_{B}}\right)^{1-n}<4.9\times10^{-11}\,,\end{eqnarray}
where $\rho_{1}=R_{1}/\kappa^{2}$. Taking $\rho_{1}=10^{-29}$ g/cm$^{3}$
as the present cosmological density and $\rho_{B}=10^{-24}$ g/cm$^{3}$
as dark and baryonic matter density in our galaxy, we obtain \begin{eqnarray}
n<5\times10^{-6}\,.\end{eqnarray}
Thus the model is very close to the $\Lambda$CDM model. The bound
on $n$ becomes even tighter if we take into account constraints
from the equivalence principle \cite{Cashinji}.

Note that other $f(R)$ models we discussed in the previous section
also need to be very close to the $\Lambda$CDM model from the LGC.
In such models the parameter $m$ behaves as $m=C(-r-1)$ as $r$
approaches $-1$ (here $C$ is a positive constant). If we demand
that the present value of $m$ is of the order of 0.1 to find a deviation
from the $\Lambda$CDM model, it is generally difficult to realize
very small values of $m$ around the region $r\approx-1$ to satisfy
the constraint (\ref{mcond}).

It is also worth mentioning that the model of Ref.~\cite{Zhang},
$f(R)=R-\lambda_{1}R_{c}\exp(-R/\lambda_{2}R_{c})$ with $\lambda_{1},\lambda_{2}>0$,
which was explicitly introduced to satisfy the LGC without fine-tuned
model parameters. When $\lambda_{1}\approx1$ the model passes the
LGC for $\lambda_{2}<10^{4}$. However, this model is not cosmologically
acceptable since it does not have a late-time accelerated attractor.
We find in fact applying the criteria set forth in Ref. \cite{AGPT}
that (i) the de Sitter point $P_{A}$ is not stable and (ii) the model
does not have an intersection point with the line $m=-r-1$ except
for the point $(r,m)=(-1,0)$, which implies that there are no additional
accelerated attractors beside the unstable de-Sitter point.

After the initial submission of this article, two viable models were
independently proposed by (i) Hu \& Sawicki \cite{Hu07} and by (ii)
Starobinsky \cite{Starobinsky07}: 
\begin{eqnarray}
 &  & {\rm (i)}~~f(R)=R-\lambda R_{c}\frac{(R/R_{c})^{2n}}{(R/R_{c})^{2n}+1}\,,\label{mo1}\\
 &  & {\rm (ii)}~~f(R)=R-\lambda R_{c}\left[1-
 \left(1+R^{2}/R_{c}^{2}\right)^{-n}\right]\,,\label{mo2}
\end{eqnarray}
where $n$, $\lambda$ and $R_{c}$ are positive constants. 
In such models the following relation holds in the region $R\gg R_{c}$
\cite{ShinjiNew}: \begin{eqnarray}
m(r)\simeq C(-r-1)^{2n+1}\,,\end{eqnarray}
 where $C=2n(2n+1)/\lambda^{2n}$. For larger $n$, $m(r)$ decreases
very rapidly as $r$ approaches $-1$ to satisfy the LGC. Moreover,
since $m(r=-2)=C$ at the de-Sitter point $P_{A}$, it is possible
to find a deviation from the $\Lambda$CDM model around the present
epoch for $C$ of the order of 0.1. The detailed analysis about these
models can be found in Refs.~\cite{Hu07,Starobinsky07,ShinjiNew,TUT},
which showed that the models can be consistent with both cosmological
and local gravity constraints for $n\ge2$. Note that the model $f(R)=R-\lambda R_{c}{\rm tanh}\,(R/R_{c})$
introduced in Ref.~\cite{ShinjiNew} is also viable, which can be
regarded as the limit $n\to\infty$ in the models (\ref{mo1}) and
(\ref{mo2}). (see also Ref.~\cite{Appleby} for a similar model).

\section{Conclusions}

We have shown that the variable $m$ that characterizes the deviation
from the $\Lambda$CDM model is constrained to be $m<{\cal O}(0.1)$
from the observational data of CMB and SNIa. We find that in general
the $f(R)$ models that are cosmologically acceptable exhibit a very
peculiar behavior of the effective equation of state $w_{{\rm DE}}$:
this crosses in fact the phantom boundary and undergoes a singularity
at a redshift of a few. If future observations will give the precise
evolution of $w_{{\rm DE}}$ in the high-redshift range, it could
be possible to detect the peculiar features of the $f(R)$ models.
This appears as an interesting way to distinguish $f(R)$ DE models
from the $\Lambda$CDM cosmology.

When we consider the local gravity constraints, we find that the deviation
parameter $m$ is required to be very much smaller than unity in the
high-density region where such experiments are carried out. In the
$f(R)$ models we studied in Sec.\,II, the parameter $m$ has asymptotic
behaviour $m\propto(-r-1)$ as $r$ approaches $-1$. If we demand
that an appreciable deviation from the $\Lambda$CDM model can be
found around the present epoch ($m(z\sim0)={\cal O}(0.1)$), we find
that it is difficult to satisfy the LGC because $m$ does not decrease
very rapidly in the region with a higher density. In such models viable
cosmological trajectories satisfying all these constraints are hardly
distinguishable from the $\Lambda$CDM model. However, the recently
proposed models (\ref{mo1}) and (\ref{mo2}) can exhibit a deviation
from the $\Lambda$CDM around the present epoch while satisfying the
LGC because the parameter $m$ decreases rapidly in the higher-density
region ($R\gg R_{c}$). It would be of interest to place observational
constraints on such models including the data of matter power spectrum,
SN Ia and weak lensing as well as LGC under the chameleon mechanism.


\section*{ACKNOWLEDGEMENTS}

We thank R.~Gannouji, W.~Hu, B.~Li, D.~Polarski and A.~Starobinsky
for useful discussions. S.\,T. is supported by JSPS (Grant No.\,30318802).

\end{document}